\documentclass[iop]{emulateapj}
\usepackage{graphicx}
\usepackage{setspace}
\usepackage{natbib}
\usepackage{color}
\usepackage{amsmath,amssymb}
\usepackage{times}

\begin{document}

\title{The dark matter halo concentration and stellar initial mass function of a CASSOWARY group}
\author{A. J. Deason\altaffilmark{1,3}, M. W. Auger\altaffilmark{2}, V. Belokurov\altaffilmark{2}, N. W. Evans\altaffilmark{2}}

\altaffiltext{1}{Department of Astronomy and Astrophysics, University
  of California Santa Cruz, Santa Cruz, CA 95064, USA; alis@ucolick.org}
\altaffiltext{2}{Institute of Astronomy, Madingley Rd, Cambridge, CB3
  0HA, UK}
\altaffiltext{3}{Hubble Fellow}
\date{\today}

\begin{abstract}

We exploit the group environment of the CAmbridge Sloan Survey Of Wide ARcs in the skY (CASSOWARY) $z=0.3$ lens J2158+0257 to measure the group dynamical mass as a complement to the central dynamical and lensing mass constraints. Follow-up spectroscopy of candidate group members is performed using VLT/FORS2. From the resulting $N=21$ confirmed members we measure the group dynamical mass by calibrating an analytic tracer mass estimator with cosmological simulations. The luminosity weighted line-of-sight velocity dispersion and the Einstein radius of the lens are used as mass probes in the inner regions of the galaxy. Combining these three observational probes allows us to independently constrain the mass and concentration of the dark matter halo, in addition to the total stellar mass of the central galaxy. We find a dark matter halo in remarkably good agreement with simulations ($\mathrm{log}_{10} M_{200}/M_\odot  = 14.2 \pm 0.2$, $c_{200} = 4.4^{+1.6}_{-1.4}$) and a stellar mass-to-light ratio which favors a Salpeter initial mass function ($\left(M/L\right)^* = 5.7 \pm 1.2$). Our measurement of a normal halo concentration suggests that there is no discrepancy between simulations and observations at the group mass scale. This is in contrast to the cluster mass scale for which a number of studies have claimed over-concentrated halos. While the halo mass is robustly determined, and the halo concentration is not significantly affected by systematics, the resulting stellar mass-to-light ratio is sensitive to the choice of stellar parameters, such as density profile and velocity anisotropy.

\end{abstract}

\section{Introduction}

The mass profiles of dark matter halos in $\Lambda$CDM cosmological simulations have a remarkably simple, universal form. The Navarro-Frenk-White (NFW, \citealt{nfw}) profile is described by two parameters: mass and concentration. This density profile can be further simplified to a one parameter family owing to the correlation between mass and concentration (typically $c \propto M^{-0.1}$). The mass-concentration relation originates from a firm theoretical prediction of the $\Lambda$CDM cosmology; lower mass halos typically form at earlier times than high mass halos when the density of the Universe was higher.

Since the seminal NFW work several theoretical studies have refined the mass-concentration relation (e.g. \citealt{bullock01}; \citealt{eke01}; \citealt{wechsler02}; \citealt{zhao03}; \citealt{neto07}; \citealt{maccio08}; \citealt{prada12}). This plethora of studies have slight differences owing to halo definitions and differing cosmologies etc., but the basic form of the profile has largely remained unchanged. The amount of work devoted to this relation reflects the fundamental nature of the prediction, and the comparison between the theoretical predictions and observations is a key test of the $\Lambda$CDM paradigm.

Observational constraints on the mass-concentration relation now span over a large mass range (dwarf galaxies to clusters), but there is some tension with theoretical predictions. For example, there is some evidence that isolated, late-type galaxies are \textit{under-concentrated} relative to $\Lambda$CDM predictions (e.g.  \citealt{kassin06}; \citealt{gnedin07}), while high mass cluster galaxies appear to be \textit{over-concentrated} (e.g. \citealt{schmidt07}; \citealt{comerford07}; \citealt{hennawi07}; \citealt{broadhurst08}; \citealt{oguri12}). In particular, cluster observations find a much steeper relation between concentration and mass, which suggests that this discrepancy will be exacerbated on the lower-mass group scale (see e.g. \citealt{fedeli12}; \citealt{oguri12}). However, the group mass scale is relatively un-explored and it is unclear how the selection biases of X-ray and lensing cluster surveys affect the observational results. Furthermore, the slope of the mass-concentration relation may not be reliably determined from fitting over a narrow mass range (see e.g., \citealt{ettori10}), whilst the significant measurement uncertainties on concentration and mass will frequently lead to an over-estimated dynamic range (i.e. the covariant scatter in the measured concentration and mass may mimic a steep relation; Auger et al.~2013).

A further complication arises due to the presence of baryonic material. Not only can the baryonic material alter the dark matter profile itself (i.e. contraction, e.g. \citealt{blumenthal86}; \citealt{gnedin04} or expansion e.g. \citealt{pontzen12}), but disentangling the contribution of the baryonic and dark matter material to the total mass profile has proved difficult. Several authors have made use of multiple mass probes over different radial ranges to try and break this degeneracy using combinations of strong/weak lensing and stellar dynamics (e.g. \citealt{sand04}; \citealt{auger10}; \citealt{newman13}) 

Often, the observed luminosity of the central galaxy has been used to fix the stellar mass component. The adopted stellar mass-to-light ratio depends on the assumed stellar initial mass function (IMF). However, in recent years, the universality of the IMF has been highly contested. In particular, massive early-type
galaxies have been claimed to have a Salpeter (e.g. \citealt{auger10}; \citealt{treu10})
or even super-Salpeter (e.g. \citealt{vandokkum10}; \citealt{spiniello11}) IMF in
apparent discord with the lighter IMF favored for less massive
galaxies (e.g. \citealt{cappellari06}), Galactic disc stars and local
globular clusters (e.g. \citealt{kroupa93}; \citealt{chabrier03}), and in spite of the environmentally independent Milky Way IMF (\citealt{bastian10}). The prospect of a non-universal IMF has profound implications for our understanding of Galaxy formation. Independent measures of the IMF, particularly at the high mass end, are thus vital in order to resolve this issue.

In this study, we use a novel approach to probe the stellar and dark matter profile of group scale halos. We make use of the over-dense environments of the  CAmbridge Sloan Survey Of Wide ARcs in the skY CASSOWARY survey (see Section \ref{sec:cass}) to complement the central dynamical and lensing mass constraints with a group dynamical mass.

The paper is arranged as follows. In Section 2 we describe the CASSOWARY survey and in Section 3 we summarize observations of the CASSOWARY lens J2158+0257. In Section 4 we describe our dynamical modeling analysis, and Section 5 outlines our maximum likelihood method which combines lensing and dynamical constraints. Section 6 describes our results, and finally we summarize our main findings in Section 7.

\section{CASSOWARY lenses}
\label{sec:cass}

\begin{figure}
  \centering
  \includegraphics[width=8.cm, height=6.5cm]{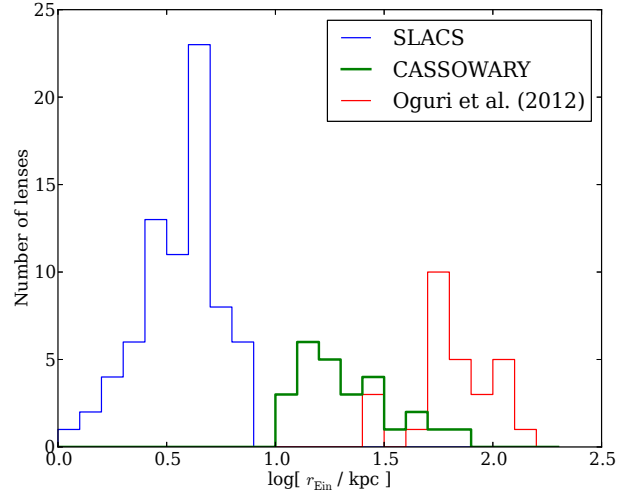}
  \caption[]{\small The distribution of Einstein radii for lenses in the SLACS (blue line), CASSOWARY (green line) and \cite{oguri12} cluster (red line) samples. The CASSOWARY lenses reside in denser environments than the SLACS field galaxies, and are typical of galaxy groups/poor clusters.}
   \label{fig:cass_slacs}
\end{figure}

The CAmbridge Sloan Survey Of Wide ARcs in the skY survey (CASSOWARY, \citealt{belokurov09}) is designed to find wide-separation ($\ge 3"$) gravitational lens systems. A typical CASSOWARY system is a massive elliptical (e.g. the Cosmic
Horseshoe, \citealt{belokurov07}) at the center of galaxy group or a
cluster (see Fig. \ref{fig:cass_slacs}) lensing a star-forming galaxy at
$1<z<3$. Compared to the The Sloan Lens ACS Survey (SLACS, see e.g. \citealt{bolton06}) lenses which are mostly field galaxies, the environment of the CASSOWARYs provides a unique opportunity to complement the central dynamical mass and lensing mass with a group dynamical mass, thus probing the mass distribution from the group center to the virial radius. 

In Fig. \ref{fig:cass_slacs} we show the distribution of Einstein radii for the CASSOWARY lenses (green line). For comparison, we show the SLACS lenses and the \cite{oguri12} sample of cluster galaxies. The CASSOWARY's are typically galaxy groups, and their Einstein radii are intermediate between field galaxies and clusters.

\section{Observations of J2158+0257}

\begin{table}
\begin{center}
\renewcommand{\tabcolsep}{0.2cm}
\renewcommand{\arraystretch}{1}
\begin{tabular}{| c  c |}
\hline
Parameter & Value\\
\hline
\multicolumn{2}{|c|}{Lens:}\\
Right ascension............................. & $21^{\rm h}58^{\rm m}43.67^{\rm s}$\\
Declination................................... & $02^\circ57'30.2"$\\
Redshift, $z_{L}$.................................. & 0.28669 \\
Magnitudes (SDSS), $g_{L}$, $r_{L}$, $i_{L}$.... & 19.38, 17.74, 17.15\\
Absolute magnitude, $M_r$.............. & -23.67\\
Effective radius............................. & 3.48''\\
Einstein radius.............................. & 3.47''\\
\multicolumn{2}{|c|}{Source:}\\
Redshift, $z_{S} $.................................. & 2.081 \\
\hline
\end{tabular}
  \caption{\small Properties of J2158+0257.}
\label{tab:central_props}
\end{center}
\end{table}

\subsection{Lens Properties}
\label{sec:lens_props}

In this study we make use of the environment of the CASSOWARY J2158+0257 lens to complement the strong lensing properties with group dynamics. The properties of J2158+0257 are summarized in Table \ref{tab:central_props}. The source redshift of the lens was measured by \cite{stark13}. In the \cite{stark13} study, the $gri$ imaging data from Sloan Digital Sky Survey (SDSS) data release 8 are used to fit the object with a singular isothermal ellipsoid lens model. The method used is described in \cite{auger11}, but extended to multiple filters and (potentially) multiple foreground light distributions. This modeling procedure fits a Sersic surface brightness profile to the foreground lensing galaxy. The resulting Einstein radius of the lens model and the effective radius of the lens are given in Table \ref{tab:central_props}.

\subsection{Galaxy Group Members}

\begin{figure*}
  \centering
  \includegraphics[width=14cm, height=7cm]{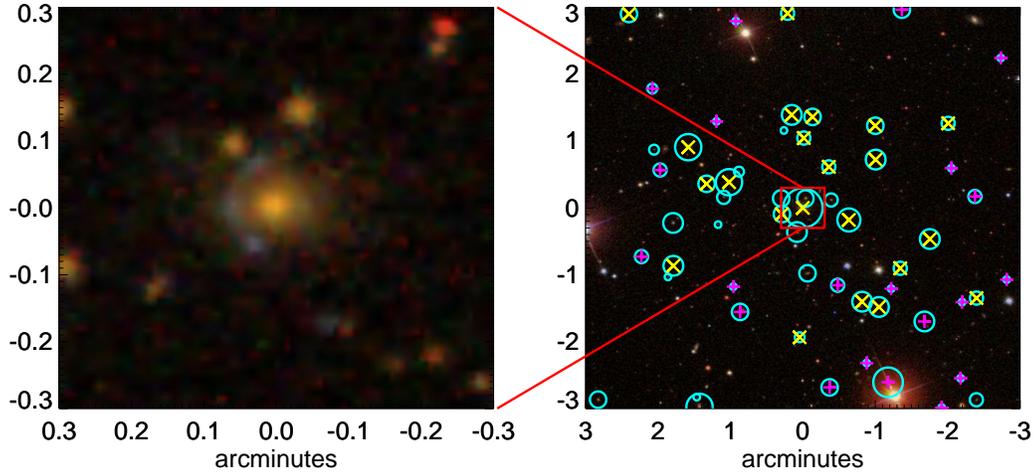}
  \caption[]{\small SDSS photometry of J2158+0257. The left-panel shows a zoom in of the central galaxy and lensed source. In the right-hand panel we show the field surrounding the lens. Cyan circles indicate candidate group members (with $|z -z_{L}| \le 0.1$). Brighter galaxies have larger symbol sizes. The yellow crosses show the (21) confirmed members from our spectroscopic survey. Candidates observed, but classified as non-members, are shown by the magenta plus signs.}
   \label{fig:lens}
\end{figure*}

Candidate group members of the J2158+0257 lens are selected using SDSS photometric redshifts. We consider relatively bright ($r_{\rm mag} < 22$) galaxies nearby the central, with similar photometric redshifts:

\begin{eqnarray}
|z_{\rm phot}-z_{L, \rm phot}| &<& 0.1 , \, \, \, \, \, \frac{\delta z}{1+z} \lesssim 0.078\nonumber\\
|\alpha-\alpha_{L}| &<& 0.05^\circ ,\nonumber\\
|\delta-\delta_{L}| &<& 0.05^\circ , \, \, \, \, \, R \lesssim 1 \mathrm{Mpc} \nonumber \\
r_{\rm mag} &<& 22 
\end{eqnarray}

Our selection criteria gives 58 candidate members. In the right-hand panel of Fig. \ref{fig:lens} we indicate the location of these possible group members. The size of the cyan symbols indicates the magnitudes of the galaxies (larger symbols for brighter objects). Targets for which we obtained follow-up spectroscopy are shown with the magenta plus symbols (non-members) and yellow crosses (members). Our follow-up spectroscopic program is outlined below.

\subsubsection{VLT Spectroscopy}

Follow-up spectroscopic observations were made using the VLT/FORS2 instrument in multi-object spectroscopy (MOS) mode. In this mode, there are 19 movable slits in a $6.8' \times 6.8'$ field-of-view (FOV). We observed three fields surrounding the central galaxy. The large FOV allows us to target several galaxies in each field; we targeted 42 (out of 58) candidate group members. Each field was centered on the central galaxy, and typical integration times were $\sim 100$ minutes per field in good conditions (seeing $\approx 1''$). Thus, we obtain a high signal-to-noise ($S/N \sim 100$) spectrum of the central galaxy in addition to lower $S/N$ spectra of the fainter candidate members sufficient to measure radial velocities. We use the 600RI+19 grism, which covers a wavelength range of $\lambda \sim 5000-8500$ \AA\ with a resolution of approximately $R \sim 1000$ at $\lambda=6780$ \AA\ . At the redshift of the J2158+0257 lens ($z \sim 0.3$), this allows us to target Ca K, Ca H, H$\beta$, Mgb and a selection of Fe lines.

\begin{figure*}
  \centering
  \includegraphics[width=15cm, height=12cm]{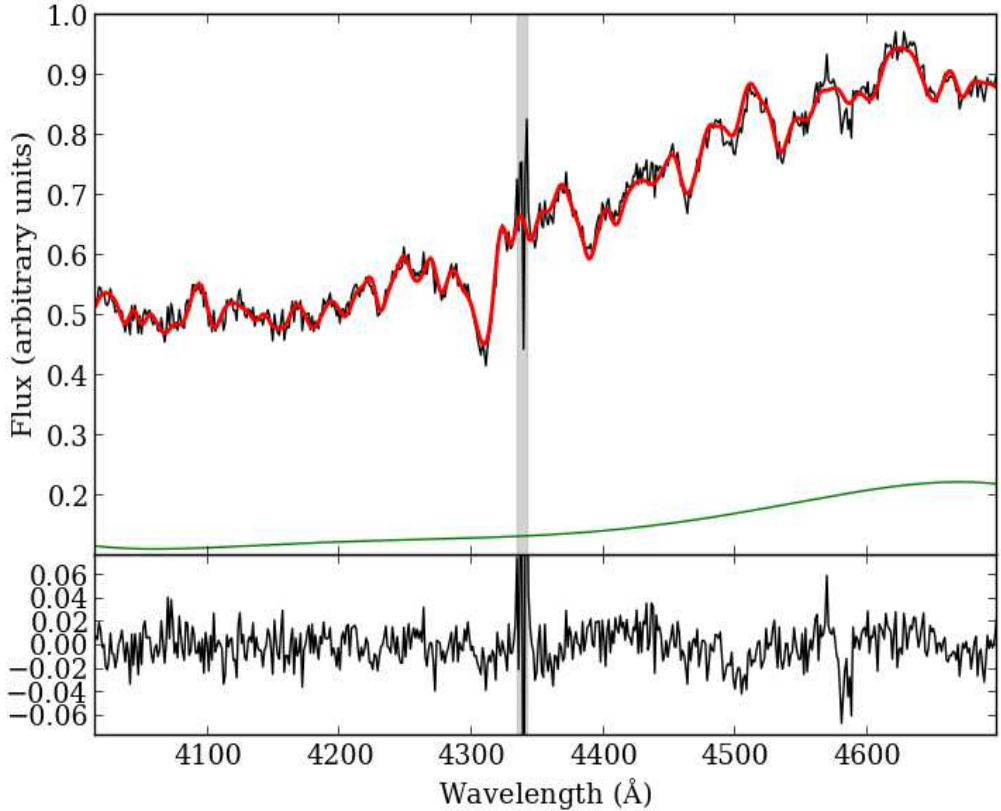}
  \caption[]{\small An example spectrum of the central galaxy J2158+0257. The solid red line shows the best-fitting template spectrum. The shaded gray region indicates the wavelengths affected by the 5577 \AA\ night sky line and the green line indicates the shape of the fitted continuum. The bottom inset panel shows the residuals of the best fit model. Note the wavelength scale is in the rest-frame.}
   \label{fig:cent_spectra}
\end{figure*}

\begin{figure*}
\centering
\begin{minipage}{0.3\linewidth}
  \includegraphics[width=6.cm, height=4.8cm]{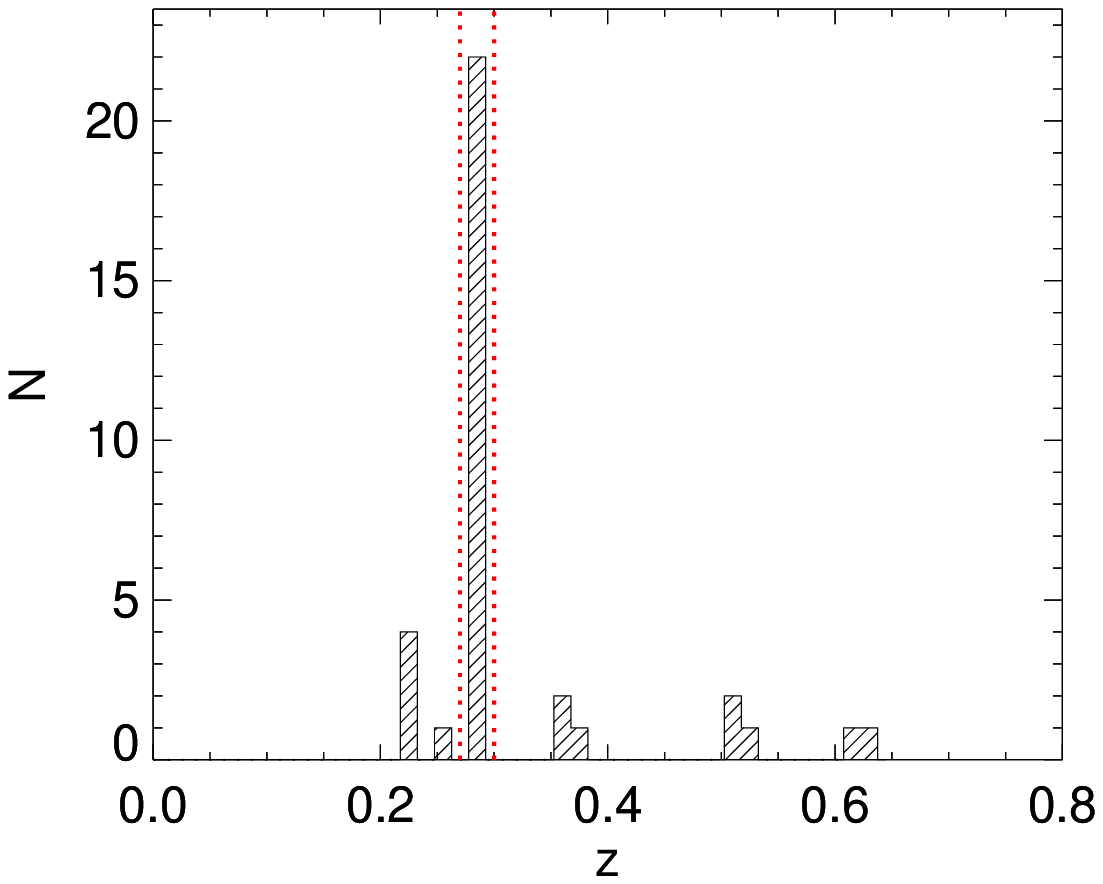}
\end{minipage}
\begin{minipage}{0.3\linewidth}
  \includegraphics[width=6.cm, height=4.8cm]{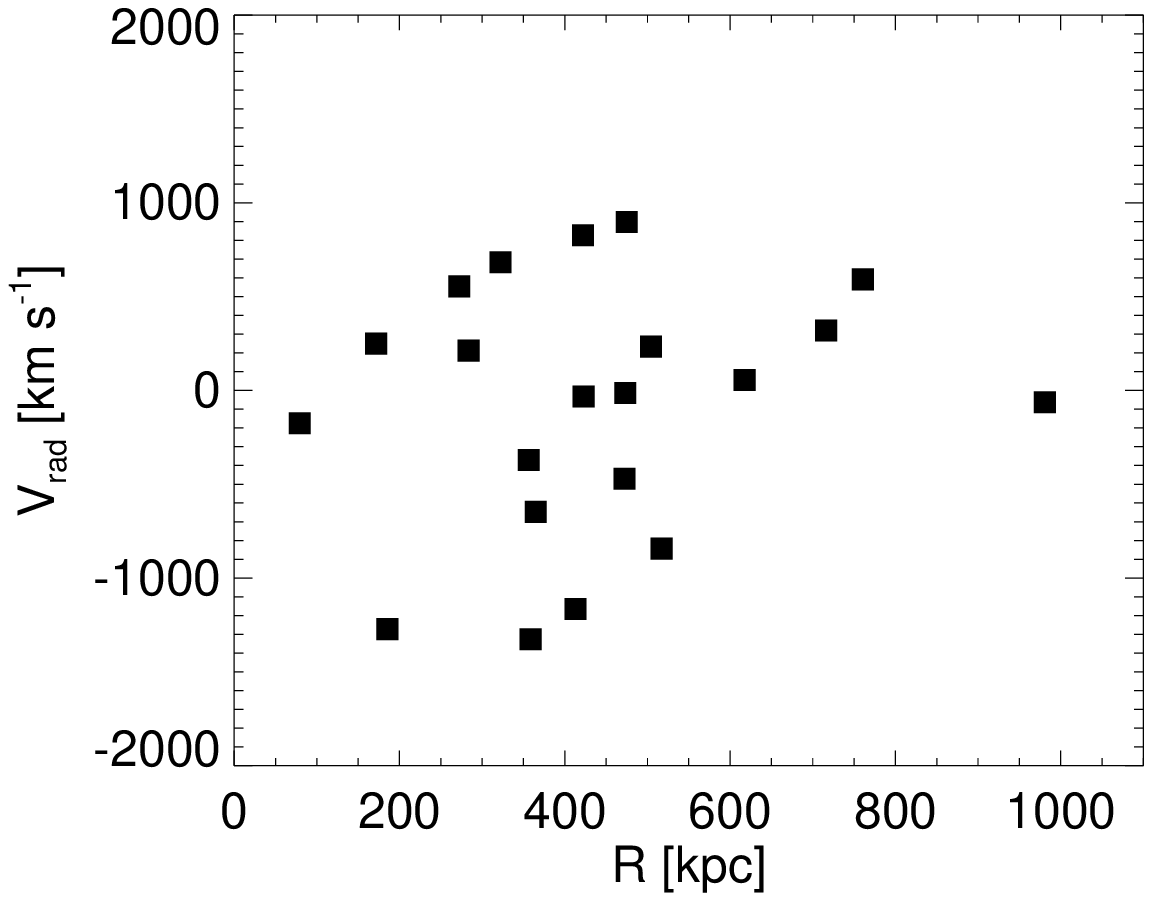}
\end{minipage}
\begin{minipage}{0.3\linewidth}
  \includegraphics[width=6.cm, height=4.8cm]{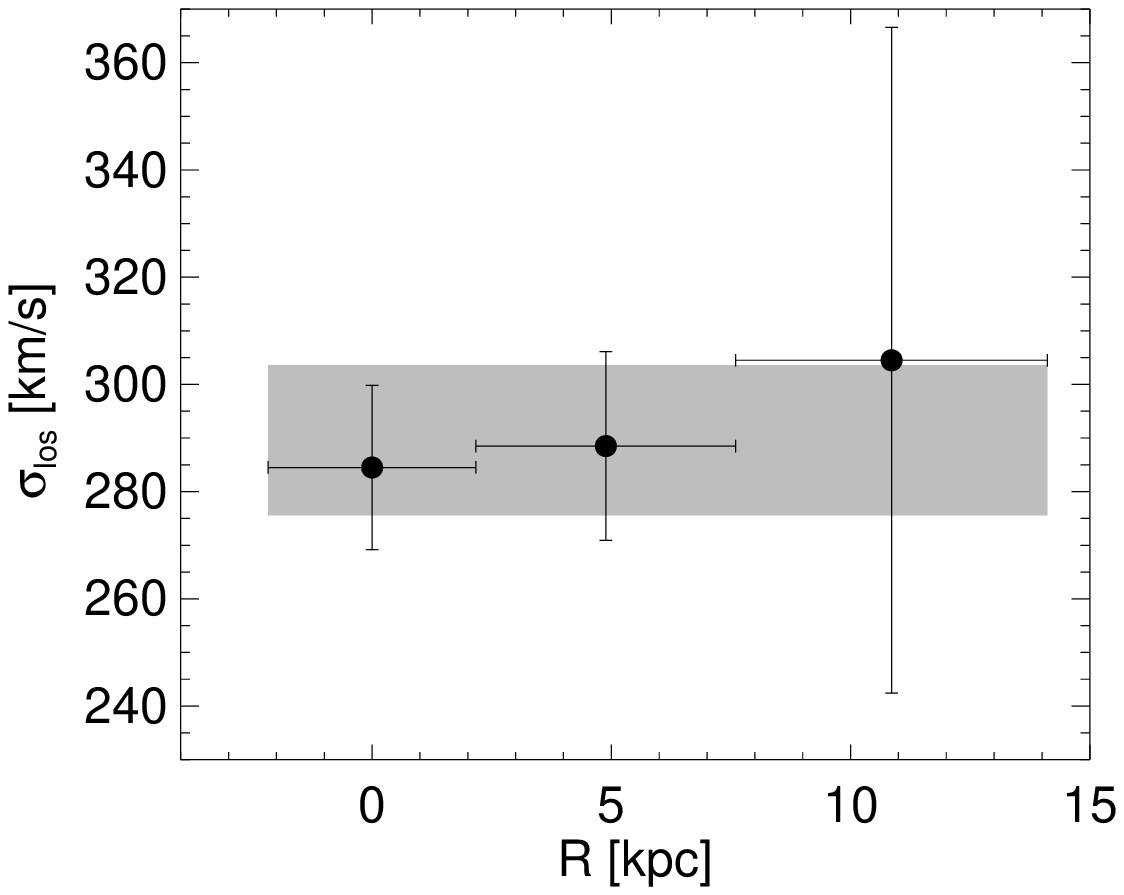}
\end{minipage}
  \caption[]{\small \textit{Left-panel:} Redshift distribution of galaxies with VLT/FORS2 spectra in the field surrounding J2158+0257. The red-dotted lines indicate the redshift range of the member galaxies. \textit{Middle-panel:} Radial velocity (relative to central galaxy) profile of group members. \textit{Right-panel:} The line-of-sight velocity dispersion profile of the central galaxy. The gray bar indicates the average velocity dispersion extracted along the central aperture.}
  \label{fig:group}
\end{figure*}

We use custom python scripts to debias, flatfield, determine the slit curvature, and wavelength calibrate the spectra using skylines. The spectra are then resampled to a constant wavelength interval, trace-straightened output frame from which the one-dimensional spectra are extracted.

A template-fitting technique is used to measure the redshifts of the candidate group members. The template spectra derive from the INDO-US stellar library, this includes a set of seven K and G giants (with a variety of temperatures) and spectra for an F2 and A0 giant. The python-based implementation is described in more detail in \cite{suyu10}. In brief, the template-fitting algorithm allows for a linear sum of template spectra to be modeled.

An example spectrum for the central galaxy is shown in Fig. \ref{fig:cent_spectra}. Here, the red line shows the best-fit template and the bottom inset panel shows the corresponding residuals. The shaded gray region indicates the masked out sky line and the green line indicates the shape of an added continuum. The high $S/N$ of the central galaxy spectrum allows us to measure the velocity dispersion (as well as the redshift). The luminosity weighted line-of-sight stellar velocity dispersion (LOSVD) profile of the central galaxy is shown in the right-hand panel of Fig. \ref{fig:group}. The black points with error bars show the velocity dispersion extracted within 3 different apertures along the $1''$ slit ($-0.5'' \, \mathrm{to}\,  0.5''$, $0.5'' \, \mathrm{to} \, 1.75''$, $1.75'' \, \mathrm{to} \, 3.25''$), and the shaded gray region is the average profile for the central aperture ($1'' \times 2.5''$).  

The left-hand panel of Fig. \ref{fig:group} shows the redshift distribution of the candidate group members. Many of these cluster around the redshift of the central galaxy ($z \sim 0.3$). We find 21 galaxies with radial velocities consistent with membership (i.e. $|V-V_{\rm cent}| < 5000$ km s$^{-1}$). In the middle panel of Fig. \ref{fig:group} we show the radial velocity profile of the group members, and their properties are given in Table \ref{tab:group_members}.

\begin{table*}
\begin{center}
\renewcommand{\tabcolsep}{0.3cm}
\renewcommand{\arraystretch}{1.}
\begin{tabular}{c  c  c  c  c  c}
\hline
ID & RA (J2000) & DEC (J2000) & $r_{\rm mag}$ & $V_{\rm rad}$ [km s$^{-1}$] & R [kpc]\\
\hline
GAL03 & 329.6990 & 2.9648 & 19.0 & 212.0 $\pm$ 6.6 & 283.6 \\
GAL04 & 329.7084 & 2.9735 & 19.0 & 897.0 $\pm$ 6.4 & 475.1 \\
GAL05 & 329.6714 & 2.9554 & 19.4 & 249.3 $\pm$ 5.9 & 171.7 \\
GAL08 & 329.6846 & 2.9816 & 19.7 & -647.7 $\pm$ 7.4 & 364.8 \\
GAL09 & 329.6528 & 2.9506 & 19.8 & -470.7 $\pm$ 5.8 & 472.3 \\
GAL10 & 329.6645 & 2.9337 & 19.9 & -14.0 $\pm$ 7.2 & 473.1 \\
GAL11 & 329.6653 & 2.9704 & 19.9 & 682.7 $\pm$ 7.7 & 322.1 \\
GAL12 & 329.7118 & 2.9440 & 20.0 & -843.4 $\pm$ 10.3 & 517.2 \\
GAL14 & 329.6684 & 2.9350 & 20.0 & -32.6 $\pm$ 10.3 & 422.8 \\
GAL18 & 329.6798 & 2.9811 & 20.1 & -370.5 $\pm$ 6.8 & 356.5 \\
GAL20 & 329.6868 & 2.9568 & 20.1 & -174.7 $\pm$ 7.9 & 79.3 \\
GAL21 & 329.6653 & 2.9789 & 20.1 & -1165.0 $\pm$ 6.5 & 412.9 \\
GAL23 & 329.7042 & 2.9644 & 20.2 & -1325.7 $\pm$ 5.9 & 358.7 \\
GAL25 & 329.7220 & 3.0067 & 20.4 & -62.9 $\pm$ 1.4 & 980.8 \\
GAL27 & 329.6486 & 2.9795 & 20.5 & 55.9 $\pm$ 10.9 & 617.6 \\
GAL31 & 329.6855 & 3.0069 & 20.6 & 591.8 $\pm$ 10.5 & 760.6 \\
GAL36 & 329.6760 & 2.9686 & 20.9 & -1272.2 $\pm$ 9.2 & 185.3 \\
GAL37 & 329.6420 & 2.9359 & 20.9 & 319.2 $\pm$ 10.8 & 716.2 \\
GAL38 & 329.6596 & 2.9433 & 20.9 & 827.1 $\pm$ 7.8 & 421.9 \\
GAL39 & 329.6818 & 2.9758 & 21.0 & 554.5 $\pm$ 7.6 & 272.3 \\
GAL41 & 329.6827 & 2.9261 & 21.3 & 233.0 $\pm$ 6.6 & 504.4 \\
\hline
\end{tabular}
  \caption{\small Confirmed member galaxies. We give the galaxy ID, right ascension, declination, r-band magnitude (from the SDSS DR8 photoZ table), and velocity and projected radius relative to the central galaxy.}
\label{tab:group_members}
\end{center}
\end{table*}

\section{Dynamical Modeling}
Our fiducial model for the total mass profile of J2158+0257 is composed of a Navarro-Frenk-White (NFW, \citealt{nfw}) dark matter halo and a Hernquist stellar bulge.

\begin{equation}
\rho_{\rm dm}(r)=\frac{M_{200}}{4 \pi r^2_s f_{200} \, r \left(1+r/r_s\right)^2},
\end{equation}
\begin{equation}
\rho_*(r)=\frac{M_* a_{\rm eff}}{2 \pi \, r \left(r+a_{\rm eff}\right)^3}.
\end{equation}
Here, $M_{200}$ is the mass within $r_{200}$, defined such that the average density within this radius is 200 times the critical density. The scale radius is $r_s=r_{200}/c_{200}$, where $c_{200}$ is the halo concentration, and $f_{200}=\mathrm{ln(}1+c_{200})-c_{200}/(1+c_{200})$. $M_*$ is the total stellar mass and $a_{\rm eff}$ is related to the effective radius of the stellar profile ($a_{\rm eff} =0.55 R_{\rm eff}$). We discuss the affect on our results if we instead adopt a Jaffe stellar profile in Section \ref{sec:beta_diss}. The effective radius and total luminosity of the stellar component are known (see Section \ref{sec:lens_props}), this leaves three free parameters in our analysis: halo mass ($M_{200}$), concentration ($c_{200}$) and stellar mass-to-light ratio ($\left(M/L\right)^*$). Here, and throughout this study, we use the r-band stellar mass-to-light ratio.

Our models correspond to everywhere positive solutions for the Jeans equations, at least within the regions of interest (cf. \citealt{an06}; \citealt{nipoti08}). At the very center ($r=0$), constant anisotropy models often become unphysical (see \citealt{an09}). Nonetheless, our investigation probes radial scales well beyond the very center, where in any case a supermassive black hole most likely resides and so the assumptions underlying our modeling already break down.

We do not include a hot gas component in our analysis. The contribution of the gas mass to the inner parts of the galaxy is negligible (see e.g. \citealt{osullivan07}), and in the outer parts the hot gas only comprises a small fraction of the total mass ($f_{\rm gas} \sim 0.1$). Since the distribution of the hot gas is similar to the halo (e.g. \citealt{allen04}), this component will be implicitly included in the dark matter halo mass ($M_{200}$). However, the uncertainties in our derived halo parameters (see Section \ref{sec:results}) are larger than $10 \%$, and thus we can safely ignore the gas component.
 
\subsection{Central Galaxy}

Our observational constraint on the central galaxy is the luminosity weighted line-of-sight velocity dispersion profile (LOSVD) out to $\sim 1R_{\rm eff}$ (see Fig. \ref{fig:group}). For given model parameters, the mass profile can be related to the LOSVD via the Jeans equations (see e.g. \citealt{mamon05} equations A7-A17). The LOSVD is then converted into the observed profile by convolving with the aperture weighted point spread function.

\begin{figure}
  \centering
  \includegraphics[width=7.5cm, height=6cm]{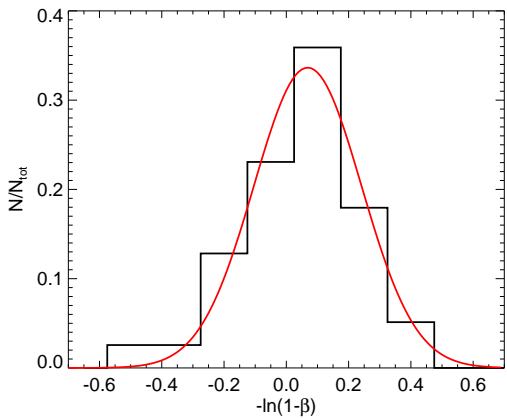}
  \caption[]{\small Measurements of velocity anisotropy in the central regions of massive elliptical galaxies from \cite{gerhard01} and \cite{cappellari07}. The red line shows a Gaussian fit to the distribution. We use this model as a prior for velocity anisotropy in our likelihood analysis.}
   \label{fig:beta}
\end{figure}

In addition to the free parameters of the mass profile ($M_{200}$, $c_{200}$ and $\left(M/L\right)^*$), the velocity anisotropy of the stars is unknown. In Fig. \ref{fig:beta} we show a distribution of measured velocity anisotropy of early type galaxies from \cite{gerhard01} and \cite{cappellari07}. These measures are generally the average anisotropy within 1$R_{\rm eff}$. Both these studies find little variation with galaxy mass. For our purposes we consider constant anisotropy models in the range $\beta = -0.5 \, \mathrm{to} \, 0.5$ with a prior distribution calibrated from these observations. The red-line shows a Gaussian fit to the observational distribution of $\beta$; we use this distribution as a prior on the adopted $\beta$ values in our maximum likelihood analysis (see Section \ref{sec:ml}).

\subsection{Group Members}

\begin{figure*}
  \centering
  \includegraphics[width=14cm, height=7cm]{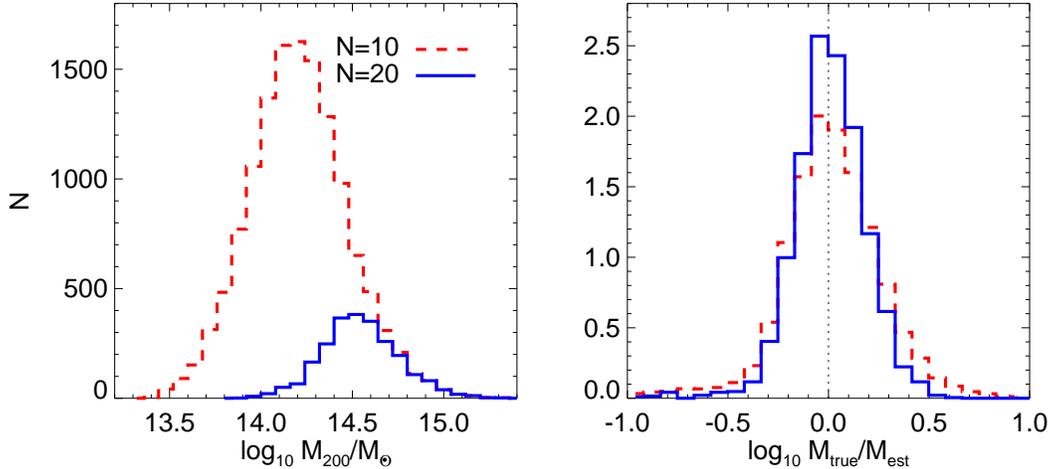}
  \caption[]{\small \textit{Left-panel:} The distribution of halo masses in the Multi-Dark simulation. The dashed-red line shows halos with 10 or more satellite galaxies, and the solid-blue line shows halos with 20 or more members. \textit{Right-panel:} The ratio of true mass to the estimated mass using a scale-free mass estimator (\citealt{watkins10}). The tracer properties (anisotropy $\beta$, density $\gamma$) and potential slope ($\alpha$) are set to the median parameters found in the simulations ($\beta \sim 0.4$, $\gamma \sim 2.0$, $\alpha \sim 0.2$). Thus, the spread in these distributions reflects the systematic uncertainties of different $\beta$, $\gamma$ and $\alpha$, as well as the statistical uncertainty associated with the number of tracers. The resulting uncertainty for $N \sim 20$ tracers calibrated from the simulations is 0.16 dex.}
   \label{fig:mest}
\end{figure*}

The observed quantities of the group members (LOS velocity and projected radius) can be used to derive a dynamical mass of the halo. We use the tracer mass estimator (TME) given in \cite{watkins10} (see also \citealt{evans03}), which assumes a spherically symmetric power-law for the halo potential $\Phi \propto r^{-\alpha}$. We use the form:

\begin{equation}
\label{eq:me}
M(<r_{\rm out})= \frac{C}{G N}\sum^N_i V^2_{\rm los, i}R^\alpha_i, \, \, \, \, C=\frac{(\alpha+\gamma-2\beta)}{I_{\alpha,\beta}} r_{\rm out}^{1-\alpha},
\end{equation}
where
\begin{equation}
I_{\alpha,\beta}=\frac{\pi^{1/2}\Gamma\left(\alpha/2+1\right)}{4\Gamma\left(\alpha/2+5/2\right)}[\alpha+3-\beta(\alpha+2)] .
\end{equation}
Here, $\alpha$ is the slope of the halo potential, $\gamma$ is the slope of the tracer density distribution, $\beta$ is the (constant) velocity anisotropy of the tracers and $r_{\rm out}$ is the 3D distance to the furthest tracer (we approximate $R_{\rm out} \approx r_{\rm out}$).

This mass estimator depends on many simplifying assumptions, such as a dynamically relaxed population in an underlying spherically symmetric dark halo with a power-law density. Furthermore, there are several (unknown) free parameters such as the tracer density slope ($\gamma$) and velocity anisotropy ($\beta$). To this end, we calibrate this mass estimator using simulations. The main assumption that this exercise relies on is that the properties of galaxy groups in the simulations is representative of ``real'' groups in the Universe, such as J2158+0257.

We use the MultiDark simulation, described in more detail in \cite{prada12}. This simulation contains about 8.6 billion particles in a (1 Gpc/$h)^3$ cube with a mass resolution of $8.7 \times 10^9 M_\odot/h$. MultiDark is larger than the Millennium simulation and uses the updated WMAP 5 cosmology. We use the $z=0.3$ halo catalog generated using the Rockstar halo finder and the Consistent Trees merger tree code (see \citealt{behroozi13a}; \citealt{behroozi13b})

In the left-hand panel of Fig. \ref{fig:mest} we show the distribution of halo masses for halos with 10 or more (dashed red line) and 20 or more (solid blue line) group members. For halos with at least 10 members, we find the average potential slope ($\alpha \sim 0.2$), group member density slope ($\gamma \sim 2.0$) and average velocity anisotropy ($\beta \sim 0.4$). These parameters are then used in the TME to measure the mass of each halo. The right-hand panel of Fig. \ref{fig:mest} shows the results of this exercise; we show the distribution of true halo mass to estimated halo mass. For halos with 10/20 or more members we only consider the 10/20 most massive tracers. We use the ``peak'' satellite mass rather than the total mass at the simulation snapshot time as this is a better proxy for stellar mass due to tidal stripping (see e.g. \citealt{reddick12}). Encouragingly the median of these distributions are centered around zero, so we have not induced any bias by using the TME. The spread of the distribution has contributions from statistical and systematic uncertainties. Thus, the dispersion accounts for varying $\alpha$, $\beta$, $\gamma$ (within the range of values in the simulations, which we assume is representative), as well as assumptions such as spherical symmetry and power-laws. 

We use the median parameters of $\alpha$, $\beta$, $\gamma$ from the simulations to estimate the dynamical mass of J2158+0257 from the group members. We adopt an uncertainty of 0.16 dex based on the spread of values for $N=20$ members in the simulations. Using equation \ref{eq:me} we find:

\begin{equation}
\mathrm{log}_{10} M(< 980 \mathrm{kpc})/M_\odot = 14.22 \pm 0.16
\end{equation}

\begin{table}
\begin{center}
\renewcommand{\tabcolsep}{0.2cm}
\renewcommand{\arraystretch}{1.}
\begin{tabular}{| c  c |}
\hline
\multicolumn{2}{|c|}{\textbf{Central Velocity Dispersion}}\\  
\hline
Radial Bin & $\sigma_{\rm los}$ [km s$^{-1}$] \\
-0.5'' to 0.5'' & $284.5 \pm 15.3$ \\
0.5''  to 1.75'' & $288.5 \pm 17.6$ \\
1.75'' to 3.25'' & $304.5 \pm 62.1$ \\
\hline
\multicolumn{2}{|c|}{\textbf{Einstein Radius}}\\  
\multicolumn{2}{|c|}{$R_{\rm ein} = 3.47" \pm 0.15 "$}\\
\hline
\multicolumn{2}{|c|}{\textbf{Group Dynamical Mass}}\\  
\multicolumn{2}{|c|}{$\mathrm{log}_{10} M (r < 980 \mathrm{kpc})/M_\odot = 14.22 \pm 0.16$}\\
\hline
\end{tabular}
  \caption{\small Summary of observational constraints from lensing and dynamics.}
\label{tab:obs_constraints}
\end{center}
\end{table}

\section{Likelihood Analysis}
\label{sec:ml}

\begin{figure*}
  \begin{minipage}{0.5\textwidth}
  \centering
  \includegraphics[width=8.5cm, height=8.5cm]{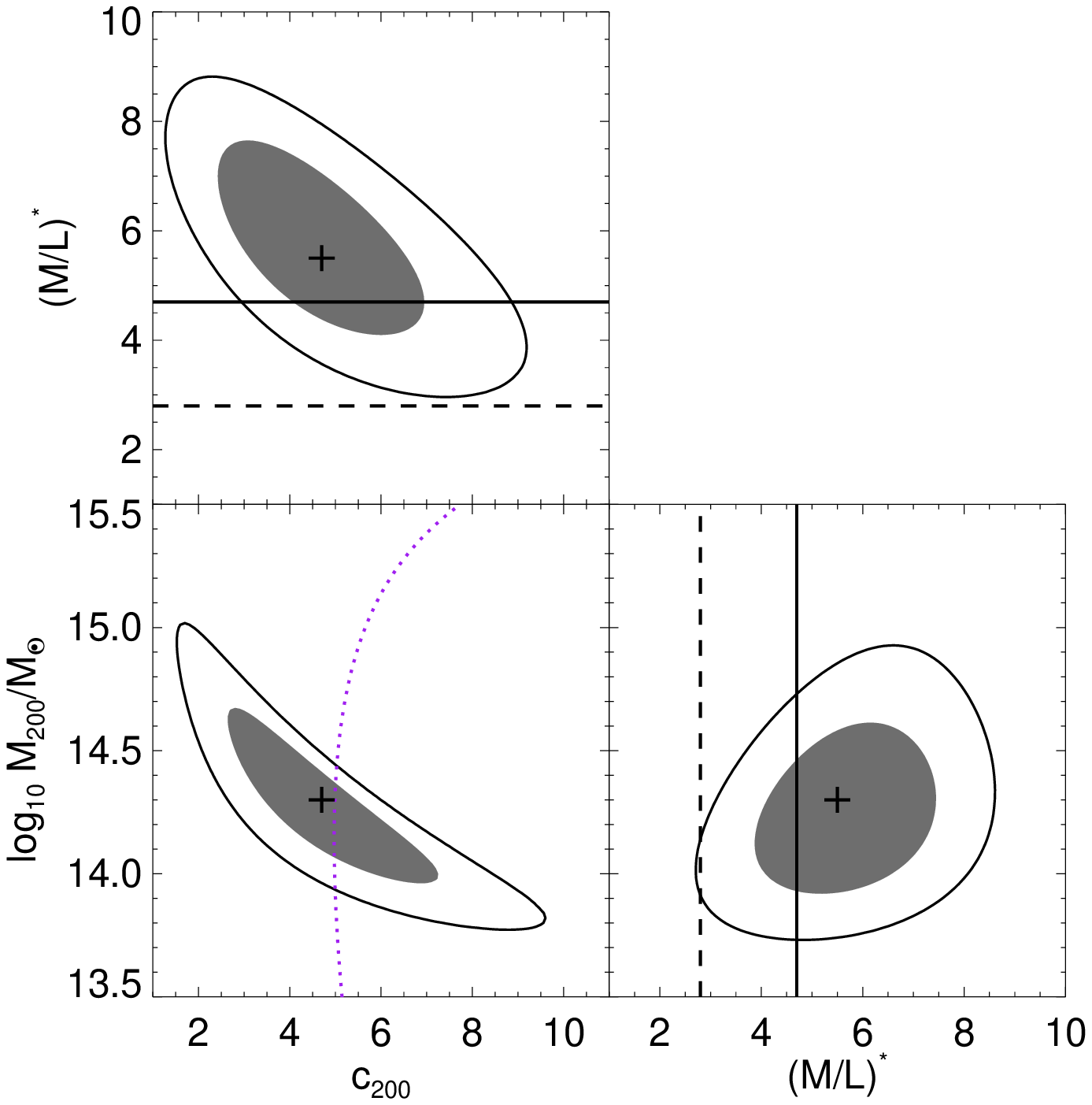}
  \end{minipage}
  \begin{minipage}{0.5\textwidth}
    \centering
    \includegraphics[width=8.5cm, height=8.5cm]{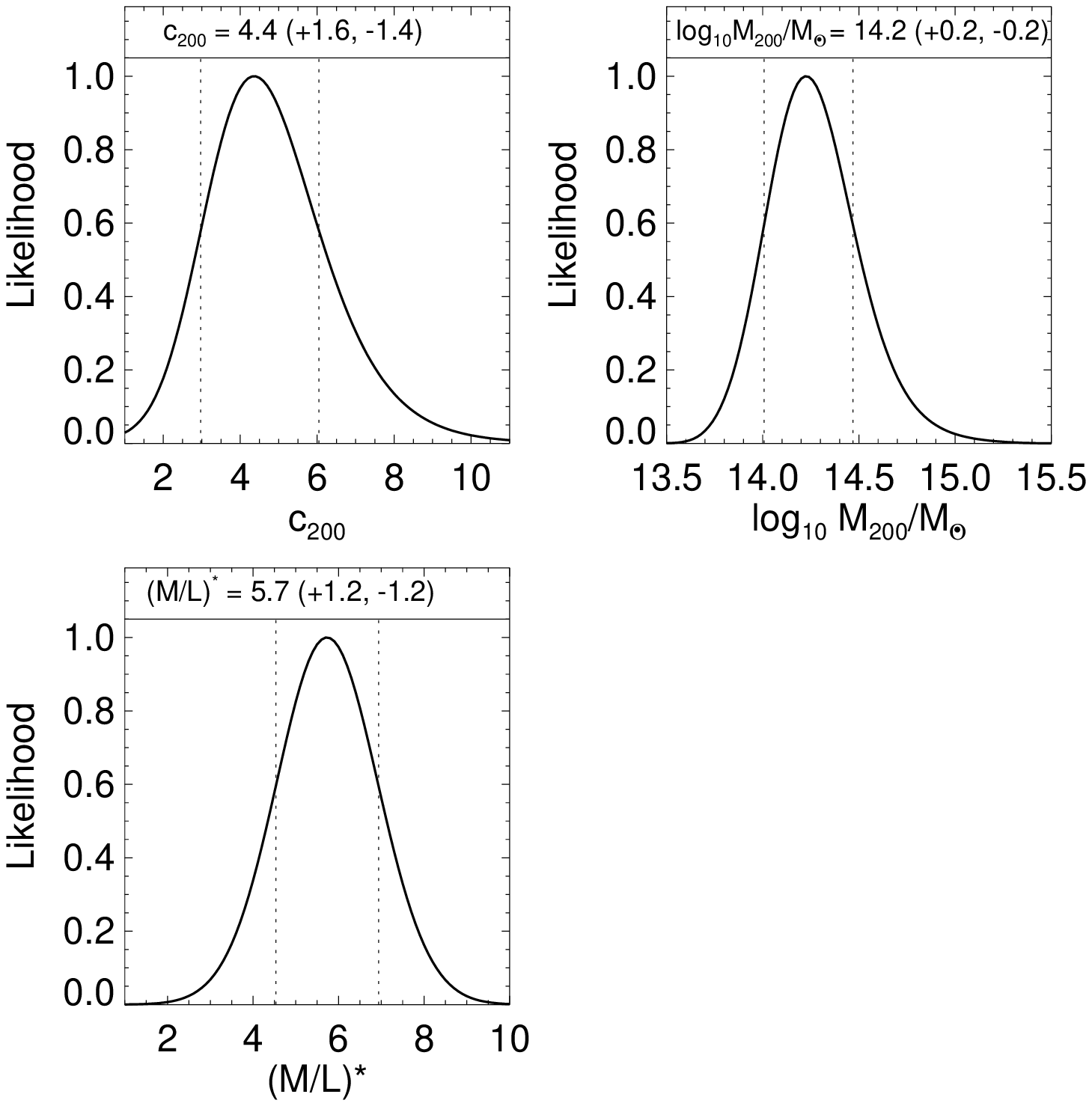}
   \end{minipage}
  \caption[]{\small \textit{Left-panel}: Likelihood results.  The gray shaded region and the solid black line indicate the 1$\sigma$ and 2$\sigma$ confidence contours respectively. We marginalize over the velocity anisotropy of the central galaxy using a prior on $\beta$ motivated from previous observational constraints (see Fig. \ref{fig:beta}). The purple dashed line shows the mass-concentration relation given by \cite{prada12}. The dashed and solid black lines indicate the approximate stellar mass-to-light ratios appropriate for Chabrier and Salpeter IMFs respectively. \textit{Right-panel}: The marginalized likelihood distributions for each parameter in our model. The dotted lines indicate the 1$\sigma$ uncertainties.}
   \label{fig:like}
\end{figure*}

A summary of the various observational constraints on the J2158+0257 group are given in Table \ref{tab:obs_constraints}. We use a maximum likelihood analysis to estimate the halo parameters $M_{200}$, $c_{200}$ and the stellar mass-to-light ratio. Our likelihood function is defined as:

\begin{eqnarray}
\label{eq:like}
\mathrm{ln} L = -\frac{1}{2}\frac{\left(R_{\rm ein, obs}-R_{\rm ein, model}\right)^2}{\sigma(R_{\rm ein, obs})}\\ \nonumber
-\frac{1}{2}\frac{\left(M_{\rm dyn, obs}-M_{\rm dyn, model}\right)^2}{\sigma(M_{\rm dyn, obs})} \\ \nonumber
-\frac{1}{2}\sum^3_k\frac{\left(\sigma_{\rm los, obs}-\sigma_{\rm los, model}\right)^2}{\sigma(\sigma_{\rm los, obs})}
\end{eqnarray}

A grid of models is used in the ranges: $\mathrm{log}_{10} M_{200}/M_\odot \in [13.5, 15.5]$, $c_{200} \in [1,12]$, $\left(M/L\right)^* \in [1,10]$. For each model family $R_{\rm ein}$, $M_{\rm dyn}$ and $\sigma_{\rm los}$ are  computed and used in equation \ref{eq:like}. In the computation of $\sigma_{\rm los}$ we use constant velocity anisotropy values in the range $\beta \in [-0.5,0.5]$, and adopt a prior on $\beta$ (see Fig. \ref{fig:beta}) to marginalize over the likelihood distributions. The affect of varying $\beta$ on our results is discussed in Section \ref{sec:beta_diss}.

\section{Results}
\label{sec:results}

\begin{table}
\begin{center}
\renewcommand{\tabcolsep}{0.2cm}
\renewcommand{\arraystretch}{1.2}
\begin{tabular}{| c  c |}
\hline
$\mathrm{log}_{10} M_{200}/M_\odot$........ & $14.2^{+0.2}_{-0.2}$\\
$c_{200}$..........................& $4.4^{+1.6}_{-1.4}$\\
$(M/L)^*$...................& $5.7^{+1.2}_{-1.2}$\\
\hline
\multicolumn{2}{|c|}{\textbf{Maximum Likelihood Model}}\\  
\multicolumn{2}{|c|}{$\sigma_{\rm los} = 280.6, 291.5, 291.8$ km s$^{-1}$} \\
\multicolumn{2}{|c|}{$R_{\rm ein} = 3.46"$}\\
\multicolumn{2}{|c|}{$\mathrm{log}_{10} M (r < 980 \mathrm{kpc})/M_\odot = 14.22$}\\
\hline
\end{tabular}
  \caption{\small Model parameters from our likelihood analysis. We also give the relevant observational quantities for the maximum likelihood model. Comparison with the values in Table \ref{tab:obs_constraints} shows that the model is an excellent fit to the data.}
\label{tab:like}
\end{center}
\end{table}

In Figure \ref{fig:like} and Table \ref{tab:like} we summarize the results of our maximum likelihood analysis. We find a halo mass of $\mathrm{log_{10}} M_{200}/M_\odot = 14.2^{+0.2}_{-0.2}$, halo concentration $c_{200}=4.4^{+1.6}_{-1.4}$ and stellar mass-to-light ratio $\left( M/L \right)^*=5.7^{+1.2}_{-1.2}$. Note that this $\left( M/L \right)^*$ corresponds to a stellar mass of $M^* \sim 1.2 \times 10^{12} M_\odot$ for the central galaxy.

The left-hand panels of Fig. \ref{fig:like} show the likelihood contours. The gray shaded region is the 1$\sigma$ confidence region, while the solid black line indicates $2\sigma$ confidence. The purple dotted line shows the mass-concentration relation derived by \cite{prada12}. The solid and dashed lines indicate the approximate r-band stellar mass-to-light ratios appropriate for Salpeter and Chabrier IMFs respectively. We use the relation between $g-r$ color and r-band stellar mass-to-light ratio derived by \cite{bell03}. This relation is corrected by -0.1 dex (Chabrier) and +0.15 dex (Salpeter) for the IMFs under consideration. Here, we have used the SDSS k-corrected and extinction corrected photometry for the central galaxy. The right-hand panels show the marginalized likelihood distributions for each free parameter. 

Our maximum likelihood model is in excellent agreement with the predictions from simulations; our best-fit halo mass and concentration fall almost exactly on the mass-concentration relation found by \cite{prada12}. The stellar mass-to-light ratio $\left(M/L \right)^* \sim 5.7$ favors a Salpeter IMF, and is $\sim 2\sigma$ away from the expectations for a Chabrier IMF (but see Section \ref{sec:beta_diss}). This is in good agreement with several authors who have argued that more massive elliptical galaxies disfavor bottom light IMFs (e.g. \citealt{auger10}; \citealt{cappellari12}).

\subsection{Stellar Velocity Anisotropy and Density Profile}
\label{sec:beta_diss}

\begin{figure}
  \centering
  \includegraphics[width=8.5cm, height=8.5cm]{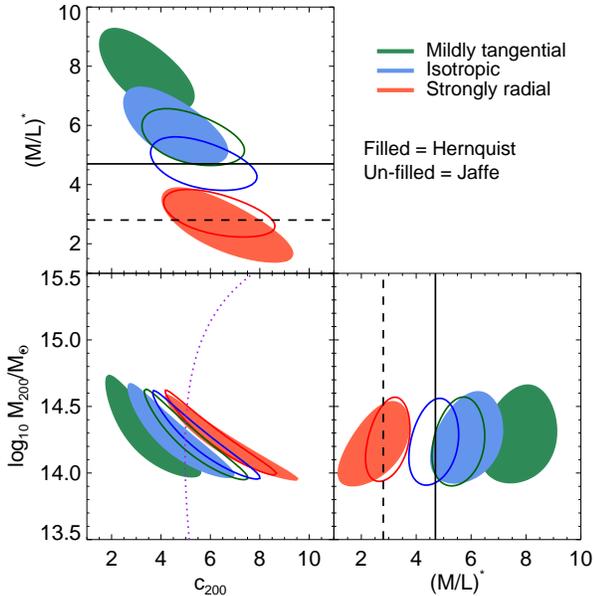}
  \caption[]{\small The effect of varying the velocity anisotropy and density profile of the central stellar distribution on our derived mass profiles. The blue, red and green filled/un-filled contours show the 1$\sigma$ confidence intervals when isotropic ($\beta=0$), strongly radial ($\beta =0.5$) and mildly tangential ($\beta = -0.5$) orbits are assumed for Hernquist/Jaffe stellar density profiles. The black solid/dashed lines indicate the approximate stellar mass-to-light ratio for a Salpeter/Chabrier IMF. There is a strong degeneracy between stellar mass-to-light ratio and velocity anisotropy.}
   \label{fig:like_aniso}
\end{figure}

One of the main assumptions in our analysis is the velocity anisotropy of the stars in the central galaxy (within $\sim 1 R_{\rm eff}$). We used a prior on $\beta$ based on previous constraints of this value in the literature. However, it is worth discussing the influence of velocity anisotropy on the derived results. In Fig \ref{fig:like_aniso} we show the 1$\sigma$ contours when we adopt $\beta=-0.5$ (green contours), $\beta=0$ (blue contours) and $\beta=0.5$ (red contours). We also show the affect of changing the stellar density profile; filled contours are for a Hernquist profile, while the un-filled contours are for a Jaffe profile.

This figure shows that there is a strong degeneracy between stellar mass-to-light ratio and velocity anisotropy. In particular, while isotropic orbits give a $\left(M/L \right)^*$ consistent with a Salpeter IMF, strongly radial orbits favor a Chabrier IMF. There is a milder degeneracy with the adopted stellar density profile; for a given velocity anisotropy, Jaffe profiles favor slightly lower $\left(M/L \right)^*$ values.

Given the importance of the form of the IMF for galaxy formation studies, it is clear that these degeneracies are important. However, we note that an extreme radial velocity anisotropy ($\beta =0.5$) is required for the models to be consistent with a Chabrier IMF. This strong radial anisotropy is 3$\sigma$ away from the median of the observed distribution of $\beta$ found by \cite{gerhard01} and \cite{cappellari07} (see Fig. \ref{fig:beta}). Moreover, a stellar population with such an extreme radial anisotropy may be radial-orbit unstable (\citealt{fridman84})\footnote{However, we note that little is known about the radial stability of stellar systems embedded in a dark matter halo.}. Thus, while we caution that the derived IMF is sensitive to the choice of stellar parameters, an extreme (and perhaps unstable?) velocity anisotropy is required for an IMF consistent with the ``Universal'' form found in the Milky Way. 

Finally, we note that there is also a slight degeneracy between velocity anisotropy and halo concentration. However, all of these models agree within their 1$\sigma$ uncertainties, and are in very good agreement with the predictions from simulations.

\subsection{A ``Normal'' Concentration Group?}

Several studies utilizing X-ray and lensing data have probed the dark matter distribution in clusters of galaxies (e.g. \citealt{schmidt07}). Commonly the halo concentrations are over-concentrated relative to theoretical predictions (e.g. \citealt{comerford07}; \citealt{hennawi07}; \citealt{broadhurst08}). While the dark matter profiles of group mass halos are relatively less explored, it is predicted that the discrepancy at this lower mass scale ($\sim 10^{14}M_\odot$) could be even more severe (see e.g. \citealt{oguri12}).

In light of these various observational constraints, it is perhaps surprising that we find such a ``normal'' concentration for the J2158+0257 group in this work. An extrapolation of the \cite{oguri12} mass-concentration relation ($c_{\rm vir} \propto M^{-0.59}_{\rm vir}$) to our mass scale ($M_{200} \sim 10^{14} M_\odot$) would predict a halo concentration of $c_{200} \sim 10$, which is $\sim 3\sigma$ more concentrated than our maximum likelihood value.

It is clear that more measures of halo concentration on the group mass scale are needed. We intend to extend this current study to several CASSOWARY lens environments in the future.

\medskip
Finally, we have not considered adiabatically contracted, or otherwise modified, halo models in this work. Often it is suggested that observations of over-concentrated halos can be explained by the contraction of dark matter halos under the influence of baryonic material in the halo centers (but see e.g. \cite{duffy10} who show that this does not alleviate the problem). Note that the ``un-contracted'' concentration of our best-fit NFW halo for J2158+0257 would be even lower if this was the case. Also, we note that a steeper inner dark matter profile adds an extra degeneracy to the derived stellar mass-to-light ratio. For example, \cite{napolitano10} find that a Kroupa IMF with adiabatic contraction or Salpeter IMF with no adiabatic contraction can both represent their data, while \cite{auger10} find a similar result for a Salpeter IMF but show that Chabrier-like IMFs are disfavored even with extreme halo contraction. These results demonstrate the intuitive trend that steeper central dark matter distributions yield lower-normalization IMFs. \cite{sonnenfeld12} explicitly parametrize the central dark matter slope (such that $\rho_{\rm DM} \propto r^{-\gamma}$ within the scale radius of the halo) and find that a slightly super-Salpeter IMF yields a profile with slope $\gamma \sim 1.5$ while a Chabrier IMF implies $\gamma \sim 2$. The same trend is seen by \cite{newman13} for cluster-sized halos but with a lower normalization. In their work, a Chabrier IMF is consistent with an NFW halo (i.e., $\gamma = 1$) but a Salpeter IMF favors a slope of $\gamma \sim 0.5$; these results suggest that moving from a Chabrier to a Salpeter IMF leads to a halo that has a shallower logarithmic density slope by $\sim 0.5$.

\section{Conclusions}
We exploit the group environment of the CASSOWARY lens J2158+0257 to complement the central dynamical mass and lensing mass with a group dynamical mass. With these three mass constraints we are able to probe the stellar and dark matter mass distribution from the group center to the virial radius. The properties of group mass halos are poorly known, and are generally unexplored relative to higher mass clusters. This study is the first step toward probing the group mass scale in more detail using the CASSOWARY lens sample. We summarize our conclusions as follows:

\begin{itemize}

\item We model the total mass distribution with a NFW halo and Hernquist stellar bulge. The resulting halo parameters ($c_{200} \sim 4.4, \, M_{200} \sim 10^{14.2} M_\odot$) are in excellent agreement with the predictions of simulations. This is contrary to observational constraints on cluster mass scales from lensing and X-rays, where systematically higher concentrations are often found.

\item Using a prior on velocity anisotropy based on observational constraints, and adopting a Hernquist stellar profile, we find that the stellar mass-to-light ratio favors a Salpeter IMF. This is in good agreement with several other studies using lensing plus dynamics. However, we caution that this result is sensitive to the choice of stellar parameters (such as anisotropy and density). 

\end{itemize} 

\section*{Acknowledgments}
AJD is currently supported by NASA through Hubble Fellowship grant HST-HF-51302.01, awarded by the Space Telescope Science Institute, which is operated by the Association of Universities for Research in Astronomy, Inc., for NASA, under contract NAS5-26555. VB acknowledges financial support from the Royal Society. We thank Peter Behroozi for making his halo catalogs publicly available, and we thank Ortwin Gerhard for kindly providing his velocity anisotropy data. AJD thanks Charlie Conroy and Risa Wechsler for useful comments and advice. We also thank an anonymous referee for useful comments.

\label{lastpage}
\bibliography{mybib}

\end{document}